\documentclass[12pt]{article}
\usepackage[dvips]{epsfig}
\newcommand{\CO}{{CompHEP}}
\newcommand{\noi}{\noindent}

\newcommand{\into}   {$\longrightarrow \:$}

\newcommand{\gentb}{$\gamma e \longrightarrow  \nu \bar{t} b \: $}

\newcommand{\ee}{$e^+e^-\: $}
\newcommand{\vtb}{$|V_{tb}| \: $}

\newcommand{\bb}{$b\,\bar{b}\: $}

\newcommand{\qq}{$q\,\bar{q}\: $}

\newcommand{\hnull}{$H^0 \: $}

\newcommand{\SQRTSEE}{$\sqrt{s_{e^+e^-}}\:$}
\newcommand{\less}{\stackrel{ <}{\sim}}

\newcommand{\Ptb}{$p_{\perp}^b\:$}

\newcommand{\Pt}{$p_\perp\:$}
\newcommand{\Ptt}{$p_{\perp}^t\:$}

\def\PL #1 #2 #3 {Phys. Lett. {\bf#1}              (#3)  #2}
\def\MPL #1 #2 #3 {Mod. Phys. Lett. {\bf#1}        (#3)  #2}
\def\IMP #1 #2 #3 {Int. Mod. Phys.  {\bf#1}        (#3)  #2}
\def\NP #1 #2 #3 {Nucl. Phys. {\bf#1}              (#3)  #2}
\def\PR #1 #2 #3 {Phys. Rev. {\bf#1}               (#3)  #2}
\def\PP #1 #2 #3 {Phys. Rep. {\bf#1}               (#3)  #2}
\def\PRL #1 #2 #3 {Phys. Rev. Lett. {\bf#1}        (#3)  #2}
\def\CPC #1 #2 #3 {Comp. Phys. Commun. {\bf#1}     (#3)  #2}
\def\ANN #1 #2 #3 {Annals of Phys. {\bf#1}         (#3)  #2}
\def\APP #1 #2 #3 {Acta Phys. Pol. {\bf#1}         (#3)  #2}
\def\ZP  #1 #2 #3 {Z. Phys. {\bf#1}                (#3)  #2}
\def\NIM  #1 #2 #3 {Nucl. Instr. and Meth. {\bf#1} (#3)  #2}
 
\newcommand{\BS}{\bigskip}

\newcommand{\lone}{$F_1^L\: $}
\newcommand{\ltwo}{$F_2^L\: $}
\newcommand{\rone}{$F_1^R\: $}
\newcommand{\rtwo}{$F_2^R\: $}

\begin{document}

\pagestyle{empty}

\noi DESY 97-045

\BS\BS

\noi March 1997
\section*{
\vspace{4cm}
\begin{center}
\LARGE{\bf
 Probing Anomalous $Wtb$ Coupling via Single Top Production 
 at TeV Energy $\gamma e$ Colliders
       }\\
\end{center}
}

\vspace{2.5cm}
\large
\begin{center}
E. Boos$^1$, A. Pukhov$^1$, M. Sachwitz$^2$ and H. J. Schreiber$^2$ \\ 
\bigskip \bigskip  
$^1$Institute of Nuclear Physics, Moscow State University, 119899,
Moscow, Russia \\
$^2$DESY-Institut f\"{u}r Hochenergiephysik, Zeuthen, FRG \\
\end{center}
\newpage
\pagestyle{plain}
\pagenumbering{arabic}

\begin{center}
\section*{Abstract}
\end{center}
Results of complete tree level calculations of the single top
production reaction \gentb  \,at the Next Linear Collider, including the
contribution of anomalous operators  to the $Wtb$ coupling are presented.
The sensitivity for probing the structure of the $Wtb$ coupling 
in a model independent way is analyzed and found to be significantly higher than for comparable measurements at the Tevatron.


\BS\BS\BS

\large

The top quark, by far the heaviest established elementary particle, is not only a further manifestation of the Standard Model (SM)\cite{sm}, it also poses new questions.
One example is the spectacular numerical coincidence between the vacuum
expectation value $v/\sqrt{2}$ = 175 GeV
and the $t$-quark mass, measured by the CDF and D0 collaborations
\cite{cdfd0} to be 175$^{+6}_{-6}$ GeV, and extracted indirectly from fits of
precision electroweak LEP data as 177$^{+7+16}_{-7-19}$ GeV \cite{ew}.
It is an open question whether or not this is due to 
fundamental physics relations or is only accidental.
The heavy $t$-quark decays electro-weakly before hadronization \cite{bigi} and therefore it could provide a first window to help understand the
nature of the electroweak symmetry breaking \cite{peccei}.
In this context, reactions involving  a light Higgs 
boson and $t$-quark production as intermediate states are 
extremely interesting.
One example is the reaction $p \bar{p} \to W^{\pm} b \bar{b} + anything,$ 
with  the two subprocesses $p \bar{p} \to W^{\pm} H^0$
($H^0 \to b \bar{b}$) and $p \bar{p} \to t b$ ($t \to W b$)
\cite{boos0}, which - 
together with several other SM diagrams - contribute
to the $W$\bb \,final state.
Another example is the reaction
 $ \gamma e \to \nu W b \bar{b}$ \cite{boos1}. 
Here, three out of 24 SM diagrams 
involve associated Higgs boson production, 

\begin{equation}
\gamma \quad e \longrightarrow \nu \quad W^- \quad H^0,
\label{eq:1}
\end{equation}

\noi and four diagrams represent single top quark production,

\begin{equation}
\gamma \quad e \longrightarrow \nu \quad \bar{t} \quad b,
\label{eq:2}
\end{equation}

\noi with subsequent decays of the Higgs boson  (\hnull \into \bb)
and the $t$-quark ($t$ \into $W b$). 

The associated Higgs production reaction (\ref{eq:1})
has a high sensitivity for probing 
anomalous $WWH$ coupling structures \cite{boos1}, 
whereas the single top reaction (\ref{eq:2}) 
is a unique tool for measuring the \vtb \ matrix element with 
very high precision \cite{boos1,boos2,jikia}.  

In this study, we consider one of the most obvious and easily
imagined scenarios in which
the $t$-quark coupling to the $W$ boson and the $b$-quark 
is altered with respect
to the SM expectations.
In order to probe such an anomalous $Wtb$ coupling 
in a model independent way, we use the effective 
Lagrangian approach \cite{gounaris1} with 
notations in the unitary gauge as given in ref. \cite{kane}.
The Lagrangian ${\cal L}$  contains 
only necessary vertices for the process (\ref{eq:2}):

\begin{eqnarray}
{\cal L} & \left.=  \frac{g}{\sqrt{2}}\right[ & 
 W_{\nu}^-\bar{b}(\gamma_{\mu}F^L_1P_- + F_1^RP_+) t \nonumber \\ 
 & & - \left.\frac{1}{2M_W} W_{\mu\nu}
\bar{b}\sigma^{\mu\nu}(F_2^LP_- + F_2^RP_+) t  \right] + {\rm h.c.} 
\label{eq:lagrangian}
\end{eqnarray}

\noi with $W_{\mu\nu} = 
D_{\mu}W_{\nu} - D_{\nu}W_{\mu}, D_{\mu} = \partial_{\mu} - i e A_{\mu}, P_{\pm} = 1/2(1 \pm \gamma_5)$ and $\sigma^{\mu\nu} = i/2(\gamma_{\mu}\gamma_{\nu} - \gamma_{\nu}\gamma_{\mu})$.
The similarity of the  $\sigma^{\mu\nu}$-connected operators with the QED anomalous magnetic moments prompts the name 
`magnetic type' for the operators and their associated vertices.  
Within the Standard Model,  
\lone = \vtb \,and \rone = $F_2^{L,R} = 0$.
Terms containing  $\partial_{\mu} W^{\mu}$ 
are omitted in the Lagrangian.
They can be recovered by applying  the quantum
equation of 
motion through operators of the original Lagrangian
\cite{gounaris1}.
We assume CP conservation with  $F_i^{L,R} = F_i^{*L,R}$.

The corresponding Feynman rules, obtained from the 
effective Lagrangian ${\cal L}$ (eq. \ref{eq:lagrangian}),
are listed in the Appendix.
These rules for the new vertices have been implemented in 
the program package \CO \,3.2 \cite{comphep}.
Effects of the anomalous couplings are simulated by varying the $F_i^{L,R}$ parameters from their SM values.
Input parameters used in the calculations were either taken  
from the Particle Data Group report 
\cite{pdg} or are as listed below:
 $m_t$ = 170 GeV, $m_b$    =  4.3 GeV, $\alpha_{EW}$ =1/128,
\vtb = 0.9984, $M_Z$    =  91.187 GeV,   
 $\sin^2\Theta_W$    =  0.23, $M_W=M_Z \cdot \cos\Theta_W$, $\Gamma_Z$=2.50 GeV
 and $\Gamma_W$=2.09 GeV.

\begin{figure*}[h!b]
\begin{center}
\mbox{\epsfxsize=17cm\epsfysize=7cm\epsffile{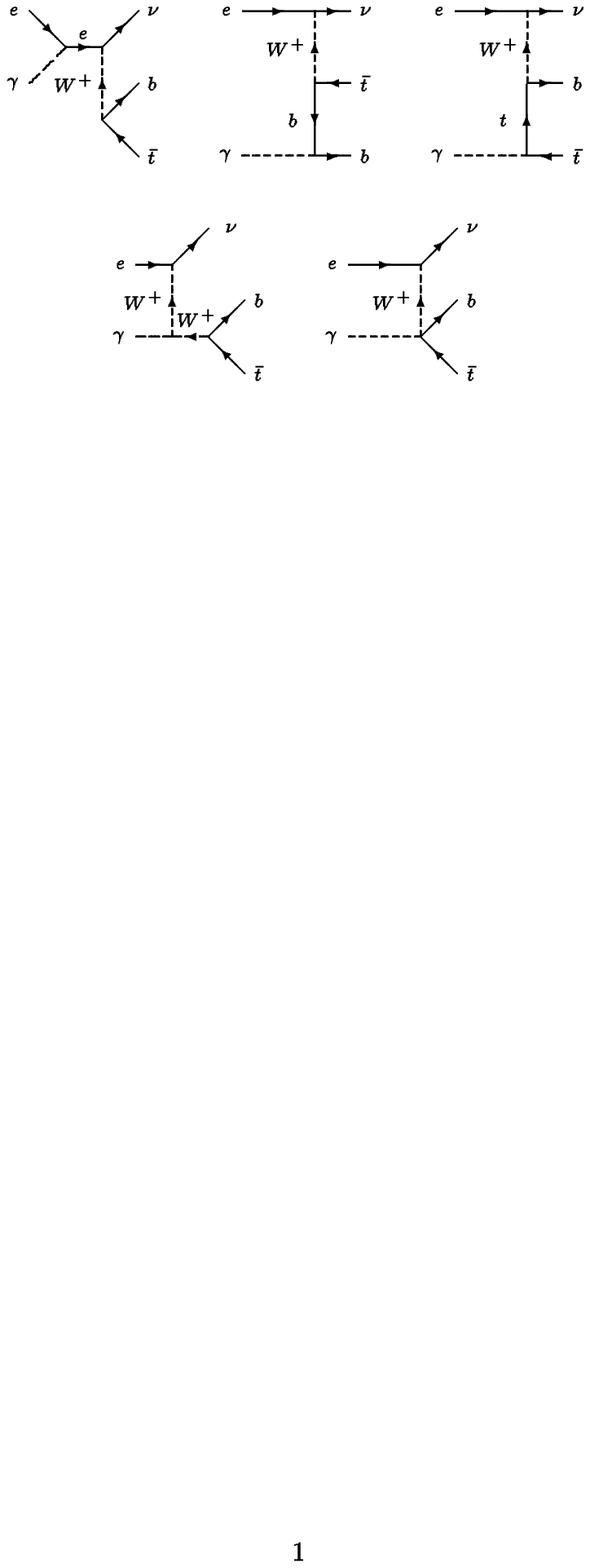}}
\end{center}
\caption{ Feynman diagrams for the reaction \gentb.}
\label{fig:feyn}
\end{figure*}

The SM tree-level diagrams contributing to the reaction \gentb
\,are shown in Fig. \ref{fig:feyn}.
The $t$-channel singularities,  occurring in the variables 
$t_{\gamma b}$ and $t_{\gamma \nu}$, have to be handled with care.  
In order to select the proper kinematic scheme for the process
considered and to smooth the singular variables we applied  special
options offered by \CO \,\cite{pukhov} (for more details see ref. \cite{boos1} ).     
We ensure  ${\cal U}(1)$ gauge invariance for the process \gentb \,by
adding the last diagram of Fig. \ref{fig:feyn} to the SM Feynman diagrams.
This non-SM diagram with the four-point $\gamma W t b$ vertex is
extracted
from the Lagrangian (3) and
contains  the sole contribution from 
the magnetic type of operators. 

The true photon beam spectrum produced by 
laser light backscattered from the incoming high energy electron beam is unknown, so we use, as 
a numerical illustration, the model-dependent photon spectrum as suggested in ref. \cite{ginzburg}. 
The convolution of the cross section for reaction (\ref{eq:2}) with this photon spectrum leaves the 
basic physical properties of the reaction unaffected but 
lowers the effective cross sections by a factor of 2-3 \cite{boos1}.

\begin{figure*}[hbtp]
\begin{center}
\mbox{\epsfig{figure=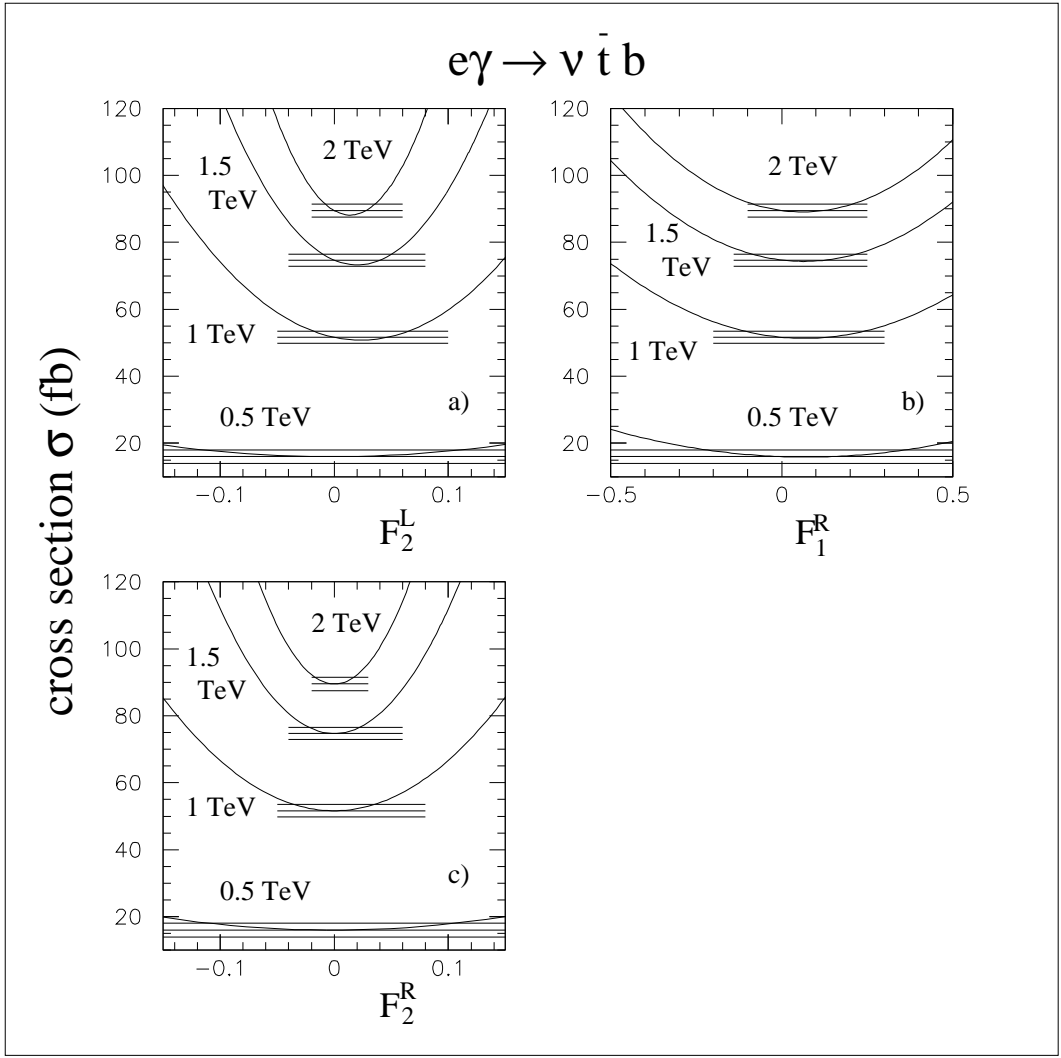,width=14cm}}
\end{center}
\caption{Cross sections of the reaction \gentb \, as functions of the
  anomalous couplings \ltwo, \rone
  \,and \rtwo at \SQRTSEE = 0.5, 1.0, 1.5 and 2.0 TeV. The horizontal 
  lines show the SM  values with the two standard deviation errors expected.}
\label{fig:2}
\end{figure*}

Fig.\ref{fig:2} shows the variation of the single top cross 
section as function  of the  
anomalous couplings  \rone, \ltwo, \rtwo, at four 
cm energies \SQRTSEE = 0.5, 1.0, 1.5 and 2.0 TeV.
Each of the figures 2a-c reflects a possible deviation of the different
anomalous couplings around zero with the other F-parameters fixed to
the SM-values.
A common feature is an increasing sensitivity with growing energies
and an enhancement of the cross section when the couplings deviate
from the SM value.
As expected from the operators' additional power of momentum  (see Appendix)
the  $F_2^{L,R}$ couplings represent a
much higher sensitivity to variations from the SM than the \rone.


With annual luminosities for the Next Linear Collider as anticipated in ref. \cite{wiik} and an  
event detection efficiency 
of 30\% for reaction (\ref{eq:2}) we  
calculate limits of the variation of \rone, \ltwo and \rtwo \,within 
two standard deviation of the SM cross section.
As can be seen from Table \ref{tab:1} the limits of the anomalous couplings 
obtained are in the interesting region \cite{peccei} of 
\begin{equation}
\frac{\sqrt{m_b m_t}}{v} \sim 0.1
\end{equation}
and do not exceed the unitary violation bounds \cite{gounaris2}
 in the one TeV scale of 
\begin{equation}
F_2^{R,L} \sim 0.8 \hspace{.5cm}{\rm and} \hspace{.5cm}  F_1^R \sim 0.6.
\end{equation}
\begin{table*}[htbp]\centering
\caption{Limits for the anomalous couplings $F_i^{L,R}$ obtained from the two standard devitation critera as described in the text  
   for annual luminosities as indicated.}
\begin{tabular}{lllll}    
                                &      &      &      &       \\
\hline\noalign{\smallskip}
                                &      &      &      &       \\
\SQRTSEE, TeV                    &  0.5 & 1.0  & 1.5 & 2.0   \\
                                &      &      &      &       \\
\hline\noalign{\smallskip}
\hspace{3cm} & \hspace{2cm} & \hspace{2cm} & \hspace{2cm} & \hspace{2cm} \\
 $\cal{L},$ fb$^{-1}$        &  50  & 200 & 300   & 500   \\
 & &  &  & \\
\hline\noalign{\smallskip}
 & &  &  & \\
 $\delta$\ltwo         &  -.1/.1 & -.020/.065 & -.01/.05 & -.008/.035 \\

 $\delta$\rtwo         &  -.1/.1 & -.035/.035 & -.022/.022 & -.016/.016 \\

 $\delta$\rone         &  -.20/.35 & -.12/.25 & -.09/.22 & -.08/.20 \\

 & &  &  & \\
\hline\noalign{\smallskip}
\end{tabular}
\label{tab:1}
\end{table*}
For comparison, recent studies of 
single top production rates including anomalous couplings  at the Tevatron 
indicate the  
bounds   -0.5 $\less F_1^R \less$ 0.5 
\cite{c-p, boos3},  
  -0.1 $\less F_2^L \less$ 0.2 and -0.2 $\less F_2^R \less$ 0.2 
\cite{boospp} which are comparable with our results  expected at NLC
energies of 0.5 TeV.
At energies above 0.5 TeV we obtain significantly higher sensitivities (see Table \ref{tab:1}).

\begin{figure*}[h!b]
\begin{center}
\mbox{\epsfig{file=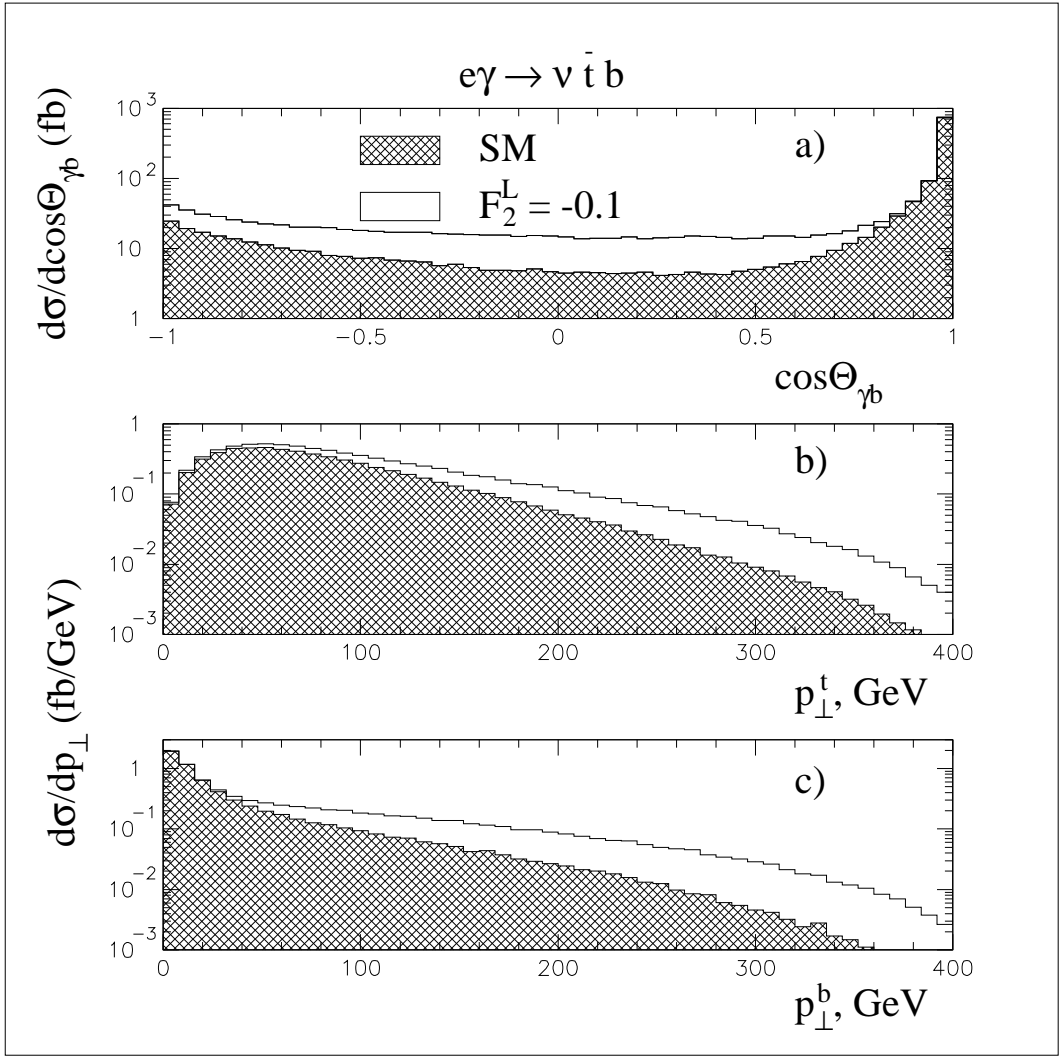,width=14cm}}
\end{center}
\caption{Cross sections of reaction \gentb \,as function of
  cos$\Theta_{\gamma b}$, 
  \Ptt \,and \Ptb at 1.0 TeV. Compared are the SM
  predictions (hatched areas) with expectations from an anomalous coupling \ltwo = -0.1.}
\label{fig:11}
\end{figure*}

The existence of anomalous couplings should also affect  the 
production properties of the final state particles of reaction (\ref{eq:2}).
As an example, Fig.\ref{fig:11}a-c show the differential cross
sections d$\sigma/$dcos$\Theta_{\gamma b}$, 
d$\sigma$/d\Ptt and  d$\sigma$/d\Ptb  
 expected for  \ltwo  = -0.1, \rtwo = \rone = 0 and \lone = SM value (open areas), compared with the SM predictions (hatched areas)\footnote{The angle $\Theta_{\gamma b}$ 
is defined as the angle of the $b$-quark with respect to the incident photon  direction in the \ee \,rest frame.}. 
In particular, the SM angular distribution d$\sigma/{\rm d}\cos \Theta_{\gamma b}$ in 
Fig. \ref{fig:11}a has a broad minimum around  $\cos \Theta_{\gamma b}
\sim$ 0 - 0.5. 
This behavior is due to the existence of the so called radiation zero
of the 2-to-2  body process \qq \into $W\gamma$   \cite{samuel}
and its time-reversed reaction 
 $\gamma W \rightarrow \bar{t} b$
 as the most
important subreaction for our consideration.
In our case, the incident $\gamma$ spectrum and the off-shell
character of the $W$-boson in addition to the contribution of the
first diagram of Fig. 1 washed out this zero to a broad minimum.
For anomalous coupling contributions the minimum becomes significantly
higher.

In the Lagrangian (3), the (V+A) operator which is 
proportional to the \rone \,coup\-ling has only an overall numerical factor
and leads to a simple shift of the  \Pt \,distributions.
On the other hand,
the new  anomalous magnetic type vertices 
(last diagram in Fig.\ref{fig:feyn}) contain an additional 
power of momentum (see Appendix) and therefore the transverse momentum distributions of the 
$t-$ and $b$-quark deviate from the SM  expectations.
As a consequence, 
such different behavior allows one to separate contributions
of the (V+A) operator from the magnetic type ones.  
Fig. \ref{fig:11}b and c show
an  excess at high \Pt for both, the  \Ptb \,and the \Ptt \,distributions. 
Clearly, cuts in the tranverse momenta and angular distributions
should lead to significantly more stringent constraints in $F_i^{R,L}$.
For illustration purpose, we require
\Ptb $>$ 40 GeV, \Ptt $>$ 80 GeV and $\Theta_{\gamma b}$ $> 10^0$ 
for the cross section calculation at 1 TeV.
Fig. 4 shows the cross section in dependence of the
coupling parameter \ltwo while
fixing the remaining $F$ parameters to their SM values. 
The bounds are improved to -0.012 $<$ \ltwo $<$ 0.058 which 
should be compared with 
 -0.020 $<$ \ltwo $<$ 0.065 (see Tab. \ref{tab:1}).
\begin{figure*}[hb!]
\begin{center}
\mbox{\epsfig{file=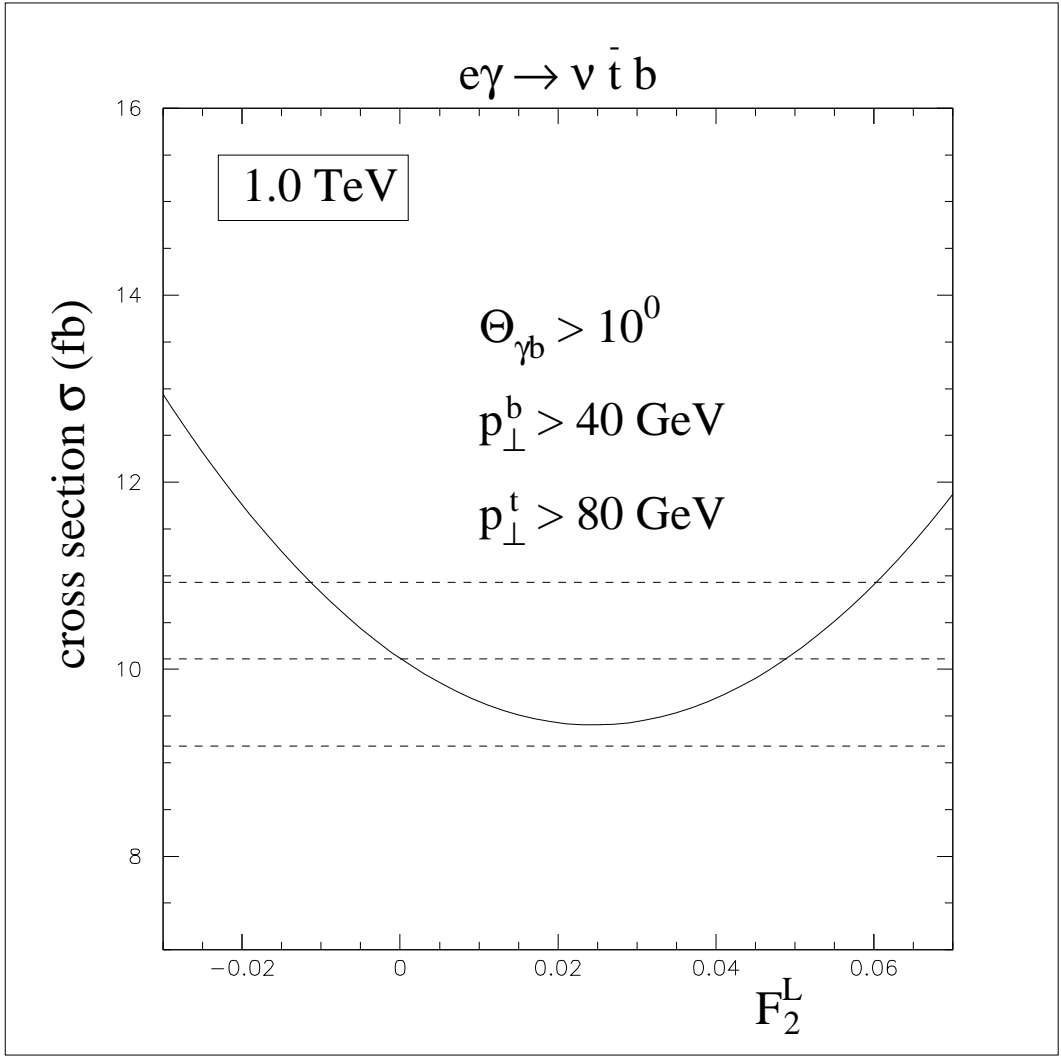,width=14cm}}
\caption{Cross section of reaction \gentb \,at 1.0 TeV in dependence of the anomalous coupling
  parameter \ltwo with angular and \Pt \,cuts as indicated. The dashed lines show the SM expectation with the two standard deviation error.}
\end{center}
\label{fig:4}
\end{figure*}

A further possibility for studying anomalous couplings could be the measurement of 
the $t$-quark partial decay width \cite{gounaris2} described by
the same effective Lagrangian ${\cal L}$. 
Here, the partial decay width is extracted from the 
single top production rate \cite{c-p} and therefore the measurement 
is not independent from the procedure given above.

A $tt$-pair production measurement would only deliver the branching ratios of the $t$-quark 
into $W$-boson and $b$-quark comparing the single and double $b$-tagging rates \cite{dan}.
Calculations show that the branching ratio is very insensitive to
variations of the F-parameters.
Even for extreme values of the parameter in the range of $\pm$1 the
branching fraction varies from 99.7\% to 99.9\%. 
Since the precision of the 
determination of the branching ratio is of the order of 10\%, a 
deviation from the SM value of 99.8\% due to the influence of 
anomalous couplings  will not be visible.

\BS


\section*{Appendix}

The Feynman rules for the vertices obtained from the   
Lagrangian (3) and implemented into the CompHEP package are as follows:

\begin{eqnarray}
\Gamma_{\mu}^{\bar{t}bW^{+}}(p,q,k) = -\left.\frac{e}{2\sqrt{2}s_{W}}
\right[ F_1^L\gamma_{\mu}(1-\gamma_5) + F_1^R\gamma_{\mu}(1+\gamma_5) \nonumber \\
\nonumber \\
-\frac{F_2^L}{2M_W}\left. (\hat{k}\gamma_{\mu}-\gamma_{\mu}\hat{k})(1-\gamma_5)
-\frac{F_2^R}{2M_W} (\hat{k}\gamma_{\mu}-\gamma_{\mu}\hat{k})(1+\gamma_5)\right]
 \nonumber
\end{eqnarray}

\BS\BS

\begin{eqnarray}
\Gamma_{\mu\nu}^{\bar{t}bW^{+}\gamma}(p,q,k,r) = 
\frac{e^2}{4\sqrt{2}s_{W}M_W}
[F_2^L(\gamma_{\mu}\gamma_{\nu}- \gamma_{\nu}\gamma_{\mu})(1-\gamma_5) 
\nonumber \\
\nonumber \\
+ F_2^R(\gamma_{\mu}\gamma_{\nu}- \gamma_{\nu}\gamma_{\mu})(1+\gamma_5)]
\nonumber
\end{eqnarray}


\section*{Acknowledgments}
E.B. and A.P. are grateful to DESY IfH Zeu\-then for the kind hospitality,
and  to P. S\"oding for his interest and support.
The work has been supported in part by the 
RFBR grants 96-02-19773a and 96-02-18635a, and by the grant 95-0-6.4-38 of 
the Center for Natural Sciences of State Committee for Higher Education
in Russia.


\end{document}